\documentclass[review,3p,times,authoryear]{elsarticle}
\usepackage{lineno,hyperref}
\usepackage[T1]{fontenc}
\usepackage{ae,aecompl}
\usepackage{float}
\usepackage{balance}
\usepackage{mathrsfs}
\usepackage{enumitem, array}
\usepackage{amssymb}
\usepackage[labelfont=bf]{caption}
\usepackage{color}
\usepackage{multirow, booktabs}
\usepackage{tabularx}
\usepackage{amsmath}
\usepackage{comment}
\usepackage{hyperref}
\usepackage{bookmark}

\usepackage{color}

\usepackage{lineno}
\usepackage[tight-spacing=true]{siunitx}

\usepackage{calc}
\usepackage{graphicx}
\usepackage{pdflscape}
\usepackage{calc}

\usepackage{caption}
\usepackage{subcaption}
\usepackage[dvipsnames, svgnames, table]{xcolor}

\usepackage{parskip}
\usepackage{amssymb}  % For checkmark symbol
\usepackage{pdflscape}
\usepackage{natbib}
\hypersetup{
	colorlinks   = true, %Colours links instead of ugly boxes
	urlcolor     = red, %Colour for external hyperlinks
	linkcolor    = blue, %Colour of internal links
	citecolor   = blue %Colour of citations, could be ``red''
}

\biboptions{authoryear}
\journal{Journal of Rock Mechanics and Geotechnical Engineering}

\begin{document}

	\begin{frontmatter}
		
		

\title{Microfluidic studies of Salt Precipitation: Influence of Brine Composition, Interfacial Tension, Flow Conditions, and Chemical Additives}
		
		
		\author[1]{Karol M. D\k{a}browski\corref{cor1}} 
		\cortext[cor1]{Corresponding author}		
		\ead{karol.dabrowski@agh.edu.pl}
		\author[2]{Mohammad Nooraiepour}
		\author[2,3]{Mohammad Masoudi}
		\author[1]{Ahsan N. Soomro}		
		\author[1]{Rafał Smulski}
		\author[1]{Jan Barbacki}
		\author[2]{Helge Hellevang}			
        \author[1]{Stanisław Nagy}	
		\address[1]{Faculty of Drilling, Oil and Gas, AGH University of Krakow, al. Mickiewicza 30, 30-059, Krakow, Poland}
		\address[2]{Department of Geosciences, University of Oslo, P.O. Box 1047 Blindern, 0316 Oslo, Norway}
		\address[3]{Applied Geoscience Department, SINTEF Industry, 7465 Trondheim, Norway}

\begin{abstract}
This study investigates the interfacial tension, fluid mobility, and crystallization behavior of various saline and additive-modified solutions in a microfluidic chip environment, simulating pore-scale processes during CO\textsubscript{2} injection. The brine compositions included NaCl solutions at different concentrations, surfactant-modified fluids, alcohol-water mixtures, and ammonia solutions. Microfluidic experiments were performed on a range of flow rates and the dynamics of CO\textsubscript{2} breakthrough, brine evaporation, and salt precipitation were analyzed.

The results show that higher NaCl concentrations accelerate crystallization and increase the final fraction of the crystal, though they also introduce spatial variability and localized precipitation. Additives such as alkylbenzene sulfonate and propan-2-ol reduce interfacial tension, promote greater mobility, and suppress salt accumulation. Ammonia-based solutions demonstrate rapid ammonia bicarbonate crystallization immediately upon CO\textsubscript{2} contact, leading to elevated water saturation and frequent chip clogging.

Despite faster brine evaporation and earlier crystal nucleation at higher CO\textsubscript{2} flow rates, no significant impact was observed on initial brine saturation or final crystal coverage. Crystal growth occurs within and outside brine pools, driven by capillary flow, with spatial distributions governed by stochastic breakthrough dynamics. The random nature of the residual brine geometry results in heterogeneous crystal patterns, which were found to be repeatable but not deterministic.

\end{abstract}
		
		
	\end{frontmatter}

	\section*{Highlights}
	\begin{itemize}
    \item Higher NaCl concentrations accelerate crystallization and increase final crystal coverage
    \item Additives such as alkylbenzene sulfonate and propan-2-ol reduce interfacial tension, enhancing fluid mobility and suppressing salt precipitation
    \item Ammonia solutions rapidly form ammonia bicarbonate crystals after CO\textsubscript{2} breakthrough, leading to water entrapment and potential flow clogging
    \item Higher CO\textsubscript{2} flow rates shorten brine evaporation time and accelerate nucleation onset
    \item CO\textsubscript{2} breakthrough paths are stochastic, resulting in random residual brine distributions
    \item Residual brine geometry governs salt crystal morphology and spatial distribution
    \item Crystal growth is fed both by local evaporation and capillary-driven brine transport between pools
    \item Final crystal fraction is insensitive to flow rate within the studied range but strongly influenced by fluid properties
	\end{itemize}

\section{Introduction}
\label{Introduction}
As global efforts intensify to limit warming to 1.5$^\circ$C under the Paris Agreement, carbon capture and storage (CCS) has become a crucial technology for mitigating anthropogenic CO\textsubscript{2} emissions, particularly in sectors such as heavy industry and power generation. CCS also allows negative emissions through bioenergy with CCS (BECCS) \citep{haszeldine2018negative,date2019global,budinis2018assessment}. CCS is a critical technology for mitigating anthropogenic CO\textsubscript{2} emissions, addressing the escalating impacts of global climate change \cite{budinis2018assessment}. Among various geological storage options, deep saline aquifers are the most promising candidate due to their vast capacity and widespread distribution \citep{bachu2015review}. These reservoirs offer a viable means to sequester large volumes of CO\textsubscript{2} in a supercritical state, leveraging their depth and pressure temperature conditions to ensure long-term containment \citep{ringrose2021storage,burnside2014review}. 

Despite its promise, CCS deployment on an industrial scale remains limited by operational and geochemical uncertainties. As global net zero targets drive industrial and government interest in large-scale CO\textsubscript{2} injection, understanding and mitigating near-wellbore phenomena, such as salt precipitation, becomes crucial to ensure injection efficiency, economic viability and long-term storage integrity \citep{kalam2021carbon}.

However, injecting supercritical CO\textsubscript{2} (scCO\textsubscript{2}) into these formations induces complex thermohydromechanical-chemical (THMC) interactions with resident brine and host rock \citep{tang2021experiment}. A critical challenge arising during CO\textsubscript{2} injection is salt precipitation in the near-wellbore region, a phenomenon driven by the displacement and evaporative drying of the formation brine \citep{miri2016review, nooraiepour2025three}. As scCO\textsubscript{2} invades the pore space, it displaces the resident brine, reducing local water saturation and concentrating dissolved salts beyond their solubility limits. This process leads to localized increases in salt concentration and subsequent salt precipitation, resulting in a reduction in porosity, impaired permeability, excessive pressure buildup, and ultimately compromising injectivity and containment integrity \citep{nooraiepour2024Damage, nooraiepour2018effect, miri2016review, masoudi2021pore}. Field-scale evidence from projects such as Snøhvit and Quest underscores these detrimental effects, revealing significant permeability losses and operational challenges related to salt precipitation \citep{Hansen2013Snohvit,ringrose2020store}. Laboratory experiments and numerical simulations on the pore and continuum scales further corroborate these findings, demonstrating that salt crystallization deteriorates porosity-permeability, alters pore geometry, and disrupts fluid flow pathways \citep{masoudi2021pore, Masoudi20249988, Nooraiepour2021SciRep}.

The dynamics of CO\textsubscript{2}-induced salt precipitation is governed by factors such as the composition of the brine, the rate of evaporation, the wettability of the substrate, and the multiphase flow regime \citep{miri2016review, nooraiepour2025three, nooraiepour2018effect, dkabrowski2025surface}. Microscale studies have elucidated different crystallization pathways, showing that crystal morphology and spatial distribution within the porous medium depend on the salinity of the brine, the kinetics of evaporation and the wettability of the substrate \citep{nooraiepour2018effect,yan2025dynamics,dkabrowski2025surface}. Capillary, viscous, and gravitational forces further influence redistribution and crystallization patterns \citep{nooraiepour2025three,zhong2025review}. However, significant knowledge gaps persist, particularly regarding multicomponent brines and stochastic nucleation behavior in heterogeneous pore structures. These gaps limit predictive modeling and optimization of injection strategies \citep{pouryousefy2016effect,lucia2015multiphase}.

To investigate these phenomena at a fundamental level, laboratory microfluidic systems have emerged as powerful tools for studying reactive transport and precipitation dynamics in porous media \citep{gerami2019microfluidics,jahanbakhsh2020review,ladd2021reactive}. These systems enable high-resolution, real-time imaging of brine evaporation and salt growth, offering unprecedented insights into pore-scale clogging mechanisms \citep{HoTsai2020,Morai2020HPMicrofluidic,LIMA2021FracDry, fazeli2019microfluidic}. By precisely controlling the dynamics of the fluid, the thermodynamic conditions, and the geometry of the pores, microfluidic platforms replicate the physicochemical processes governing CO\textsubscript{2}-induced salt precipitation \citep{kim2013aquifer,nooraiepour2018salt}. Previous studies utilizing these systems have illuminated the capillary-driven film flow, multiphase interactions, and wettability effects on crystallization kinetics \citep{Ott2021Salt,dkabrowski2025surface}. Key operational parameters, including CO\textsubscript{2} injection rate, pressure-temperature conditions, brine salinity, pore geometry, and heterogeneity, have been systematically analyzed in experimental studies to assess their influence on the dynamics of salt precipitation \citep{nooraiepour2018effect, nooraiepour2025three}. Pore-scale efforts have demonstrated the probabilistic nature of mineral nucleation, affecting evaporation-precipitation on primary and secondary solid substrates, highlighting its dependence on stochastic variations in the spatiotemporal domain \citep{Nooraiepour2021SciRep,Nooraiepour2021Omega,masoudi2022effect}. These works provided crucial insights into not only the quantity of precipitated crystals but also the spatial distribution of crystalline deposits and their subsequent impact on changes in porosity and permeability in two-dimensional microfluidic systems.

 Most prior work has focused on single-component brine, primarily NaCl systems, under idealized conditions. Natural formation waters are much more complex, containing various salts such as CaCl\textsubscript{2}, MgCl\textsubscript{2}, Na\textsubscript{2}SO\textsubscript{4}, and KCl \citep{peter2022review}. These composite brines affect crystallization pathways, wettability, and porosity-permeability relationships in ways not captured by studies with NaCl alone \citep{zhang2019study,abbasi2022mixed,darkwah2024comprehensive}. Field observations show that such salts can exacerbate near-wellbore clogging and contribute to complex precipitation patterns \citep{kaszuba2005experimental,guyant2015salt,darkwah2024comprehensive,muller2011supercritical}. A 2023 review in Fuel further emphasized the need to study these multicomponent systems experimentally \citep{cui2023review}.

Despite extensive efforts, a comprehensive understanding of the dynamics of multicomponent salt precipitation remains elusive, particularly under realistic geological and operational conditions. The interplay between complex brine chemistries, heterogeneous pore structures, and dynamic flow regimes is not fully captured in current models or experiments, thereby limiting the reliability of predictive tools for field-scale CCS operations.

To address these gaps, this study presents a novel integration of realistic multicomponent brine systems and additive-modulated interfacial dynamics within microfluidic environments, an approach that has not been comprehensively explored in the prior literature. 

This work incorporates realistic brine chemistries and chemical additives in both laminar and turbulent flow regimes. It offers a comprehensive understanding of the spatiotemporal evolution of salt precipitation and its influence on porosity-permeability relationships. Modification of the interfacial tension (IFT) is performed by the additive alkylbenzene sulfonate surfactant, which shows the impact of fluid mobility on residual brine saturation.  The addition of KOH changes the pH of the solution and serves as additional nucleation promoters. The alcohol solution in addition to modification of the volatility of the IFT change fluid enhances the evaporation process. In the work, we also considered the ammonia + water solution, which modifies IFT, volatility, and also reacts upon contact with CO\textsubscript{2}. This consideration may be in particular of interest in terms of ammonia storage. 
 
That modulations of brine properties and precipitation dynamics identify potential strategies to mitigate injectivity losses. High-resolution imaging of crystal growth patterns and spatial distributions enables the development of probabilistic nucleation and reactive transport models tailored to predict salt precipitation in saline aquifers. Ultimately, this research informs optimized injection strategies to improve the scalability, safety, and cost-effectiveness of CCS projects in deep saline aquifers. In addition, the insights gained can be integrated into enhanced reactive transport models, forming field-scale injection designs and contributing to the development of predictive frameworks and risk mitigation strategies essential for the widespread deployment of CCS technologies. These findings also serve as a foundational data set for calibrating and validating multiphase reactive transport simulators, ultimately enhancing the fidelity of large-scale reservoir models used in CCS planning and risk assessment. Beyond CCS, the insights derived from this study have implications for broader subsurface engineering challenges, geothermal energy extraction, and contaminant remediation, where multiphase flow and reactive transport play critical roles.

\section{Materials and Methods}
\subsection{Microfluidic chips and laboratory setup}
This study utilizes a hydrophilic two-dimensional (2D) micromodel with rock-like pore geometries. The micromodel, measuring 20 mm (length) × 10 mm (width), features a pore size distribution ranging from 100 µm to 600 µm, with a median of 250 µm. The chip’s porosity is 0.47 ± 0.01, and its permeability is 7.2 ± 1.1 darcy.

	\begin{figure}
		\centering	
		\includegraphics[width=0.9\textwidth]{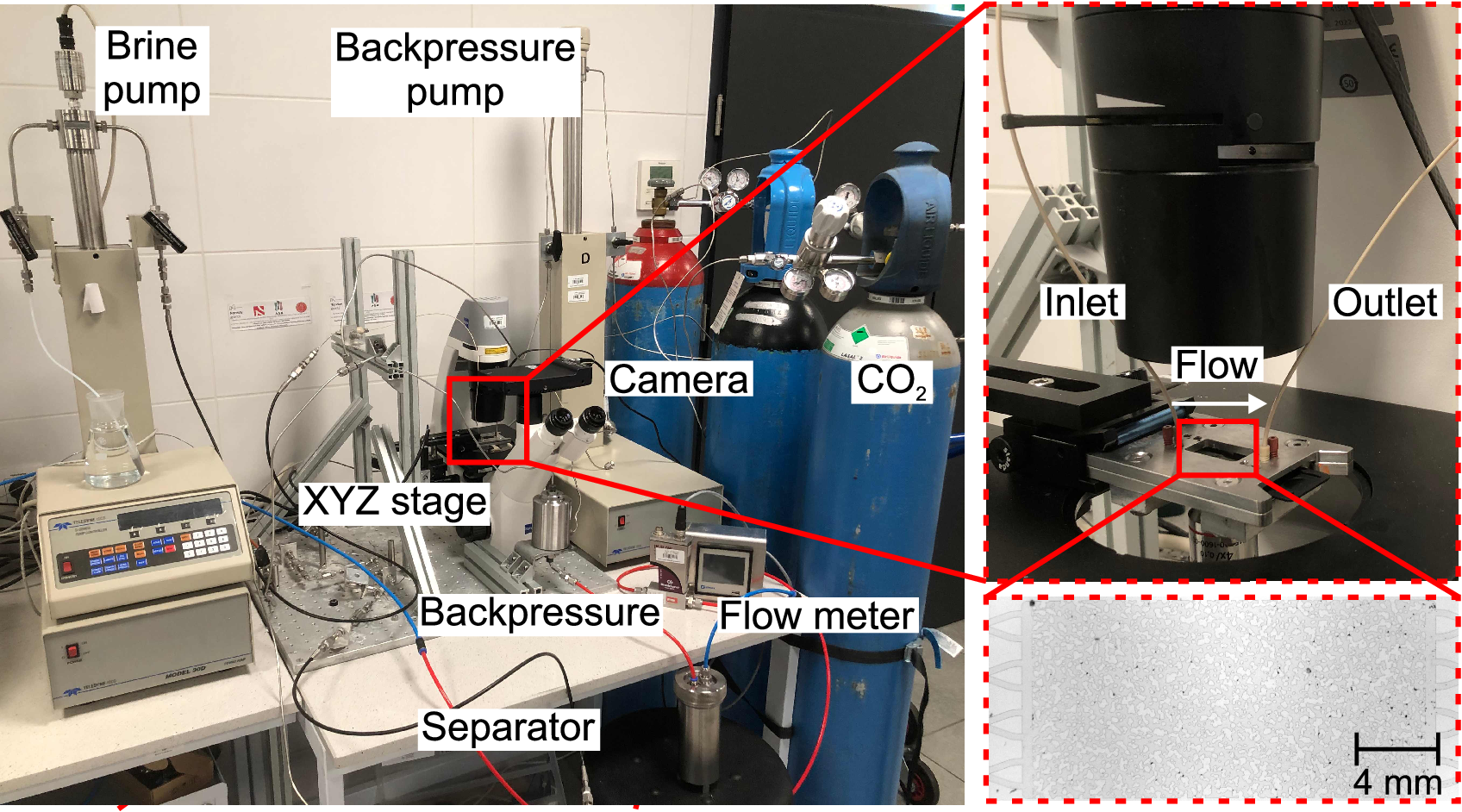}
		\caption{Photograph of the laboratory setup for microfluidic observation of brine evaporation and salt crystallization. The system includes syringe pumps for brine injection, a backpressure pump for flow regulation, a CO\textsubscript{2} cylinder, a high-speed microscopic camera, a mass flow meter, and a separator. The first inset details the chip holder, XYZ motorized stage, and flexible PEEK tubing connections, which enable smooth stage movement at 5 MPa inlet pressure. The second inset shows a microscopic image of the entire chip structure (field area: 22 × 10 mm). This setup is an enhanced version of that described by \citep{dkabrowski2025surface}.}
		\label{fig:figure1}
	\end{figure}

The evaporation-precipitation dynamics within the pore network was investigated using a custom-designed experimental setup, adapted from the system described by \citep{dkabrowski2025surface}. Modifications include enhanced flow control and the addition of spatially resolved pressure measurements along the chip. A detailed illustration of the setup, including the chip holder, the tubing, and the pore network, is provided in Figure~\ref{fig:figure1}. The imaging was carried out using a FASTEC IL5 high-speed camera equipped with a 2560 × 2080 12-bit CMOS sensor (5 µm pixel size, low noise), capable of recording at 20,000 frames per second (fps). The camera was coupled to a ZEISS Primo Vert microscope to monitor a 2 × 2 mm chip section. To capture the entire 20mm length of the chip, an XYZ motorized stage (Standa Ltd.) with a positional resolution of 0.125 µm was used for panoramic imaging, allowing spatially resolved analysis over a 2 × 20 mm domain.

Fluid injection was managed using 1/16 inch flexible PEEK tubing connected to high-precision syringe pumps (Teledyne ISCO 30D and 100DX). Gaseous CO\textsubscript{2} was maintained at 50 bar using pressure regulators, with flow rates precisely controlled by a backpressure regulator (Vinci Technologies BPR Series 10000) interfaced with the syringe pump system. Flow rates were monitored with a Coriolis flow meter (Bronkhorst IN-FLOW), while differential pressure transducers (Keller PD-33) measured pressure gradients across the chip. This configuration enabled a systematic evaluation of the effects of flow rate on crystallization dynamics. For each experiment, the input pressure was maintained at a constant 50 bar (5 MPa), while the outlet pressure decreased with increasing flow rate, reflecting injection scenarios under constant pressure injection conditions.

Interfacial tension (IFT) and contact angle measurements were performed using an IFT Cell 1000 from Vinci Technologies. A tilt base accessory allowed one to determine both the advancing and receding contact angles using a method described in \citep{pierce2008understanding}. The micromodel exhibited a static contact angle of 27$^\circ$, indicating hydrophilic conditions.

\subsection{Fluid Compositions and Properties}
Most previous studies investigated the kinetics of crystallization using a NaCl solution brine. These studies focused primarily on the effects of concentration. This study expands beyond that scope by examining a diverse range of fluid compositions to assess their influence on salt precipitation dynamics under CO\textsubscript{2} injection conditions.

Table~\ref{tab:IFT_Flow} summarizes the interfacial tension (IFT), contact angle (receding and advancing), and flow rate conditions for eight fluid compositions. IFT values are reported at ambient pressures (0.1 MPa) and elevated pressures (5 MPa) to highlight the effect of CO\textsubscript{2} pressurization, which consistently reduces IFT due to enhanced gas-liquid interactions. Contact angles, measured at 5 MPa, reflect variations in wettability that are critical to fluid distribution and precipitation behavior. 

The first three fluids serve as reference solutions. These are NaCl solutions in water with molalities of 1, 2.5 and 5 mol/kg (marked with IDs W1, W2.5 and W5). These solutions are benchmarks for tests that investigate the impact of additives. This approach enables the determination of the influence of concentration on the crystallization process. NaCl concentration emerges as a primary driver of the crystallization kinetics. Higher concentrations (e.g., 5 mol / kg in W5) increase the IFT, thus reducing fluid mobility within the micromodel and accelerating brine drying, which shortens crystallization onset times and increases the final crystal fraction. 

To demonstrate the behavior of the reservoir fluid, a natural brine sample taken from the Stargard Szczecinski GT-2 geothermal well in northwest Poland is studied. This brine is marked with the ID of S2.4, as the salt concentration is 2.4 mol/kg in the solution. Although it has lower salinity than the 2.5 mol/kg solution (ID W2.5), this natural sample exhibits higher IFT values. 

The first additive examined is a 0.2$\%$ KOH in a 5 mol/kg NaCl brine solution. This additive forms a base solution with high pH. Therefore, it is marked with the ID of B5. During CO$_2$ flow, CO$_2$ dissolves in water and forms carbonic acid. The addition of KOH helps maintain a constantly high pH. At high pressure, KOH does not affect the IFT or the contact angle.

In contrast, the addition of $0.1\%$ alkylbenzene sulfonate (ABS) to a brine of 5 mol/kg NaCl (ID ABS5) significantly decreases IFT, which increases fluid mobility. The addition of ABS also increases the contact angle. This more hydrophobic system is expected to exhibit a lower crystal coverage \cite{dkabrowski2025surface}. It should be pointed out that high salinity reduces ABS solubility in brine, thereby limiting the final IFT reduction.

Similarly, mol/kg NaCl brine with propan-2-ol (isopropyl alcohol) decrees IFT but also decreases contact angle. Therefore, this solution is more hydrophilic than the pure NaCl brine. Isopropyl alcohol has a higher vapor pressure than brine (5800 Pa vs. 3200 Pa) and therefore exhibits a higher evaporation rate, which is expected to increase the crystallization rate. Isopropyl alcohol does not dissolve NaCl and, when mixed with water, decreases the total solubility of NaCl, which can also affect the timing of crystal formation. The choice of a 50$\%$ alcohol concentration allows the maximum impact of alcohol while maintaining adequate NaCl solubility. 

Finally, the mixture of 75$\%$ NaCl brine + 25$\%$ ammonia (ID A4) decreases both the IFT and the constant angle. Ammonia has a 20 kPa vapor pressure and is therefore expected to enhance crystal formation. Our primary interest in studying ammonia stems from the theoretical possibility of energy storage (in the form of ammonia) in salt caverns. In salt caverns, permeable rock interlayers can be found between the salt. These interlayers can be affected by storage leakage when ammonia and brine migrate through permeable rock. Therefore, measuring the interaction between brine, ammonia, CO$_2$, and rock can provide valuable information for storage safety and efficiency. 25$\%$ ammonia solution is selected because it represents the highest concentration of aqueous ammonia solution available.  

\begin{table}
    \centering
    \caption{\textbf{Interfacial Tension (IFT), Contact Angles receding (R) advancing (A) , and Flow Rate Conditions for Experimental Fluids}}
    \label{tab:IFT_Flow}
    \footnotesize    
    \begin{tabular}{l c c c c c c c c c c}
        \toprule
        Composition & NaCl  & Sample  & IFT (mN/m) & IFT (mN/m) & R  & A & \multicolumn{4}{c}{Flow Rate (mL/min)} \\
        & (mol/kg) & ID & $p=0.1$ MPa & $p=5$ MPa & ($^\circ$) & ($^\circ$) & 50 & 150 & 300 & 1000 \\
        \midrule
       NaCl brine & 1 & W1 & 72.9 & 41.8 & 43 & 53 & &  &  & \checkmark \\
       NaCl brine & 2.5 & W2.5 & 78.5 & 44.6 & 29 & 39 &  & \checkmark & \checkmark & \checkmark \\
       NaCl brine & 5 & W5 & 84.3 & 50.0 & 31 & 33 & \checkmark & \checkmark & \checkmark & \checkmark \\
        Stargard brine & 2.4 & S2.4 & 79.5 & 46.6 & 44 & 49 & & \checkmark & \checkmark & \checkmark \\
       NaCl brine + 0.2\% KOH & 5 & B5 & 79.1 & 50.9 & 30 & 33 &  & \checkmark & \checkmark & \checkmark \\        
        NaCl brine + 0.1\% ABS$^*$ & 5 & ABS5 & 20.8 & 10.9 & 38 & 44 & & \checkmark & \checkmark & \checkmark \\                
        50\% NaCl brine + 50\% propan-2-ol & 2.8 & I2.8 & 26.3 & 9.8 & 27 & 32 & & \checkmark & \checkmark & \checkmark \\
        75\% NaCl brine + 25\% Ammonia & 4 & A4 & 66.4 & 48.7 & 35 & 42 & & \checkmark &  & \checkmark \\
        \bottomrule
        \multicolumn{11}{l}{$^*$Alkylbenzene sulfonate}
    \end{tabular}
\end{table}

The experiments were carried out at flow rates of 150, 300 and 1000 ml / min (measured at ambient pressure after the separator) for all sample. For selected samples (W5, B5) also at 50 ml / min. This corresponds to Reynolds number 250, 800, 1600 and 8000 for each flow rate. This covers both laminar and turbulent flow regimes. Each experimental condition was repeated three times, resulting in a total of 76 laboratory reactive transport experiments. Crystallization in W1 (1 mol/kg NaCl) at 150 ml / min occurred in only one of three trials, suggesting a threshold effect related to low salinity and moderate flow.
    
\subsection{Pore-scale evaporation-precipitation dynamics and image analysis}

\begin{figure}
\centering
\includegraphics[width=0.9\textwidth]{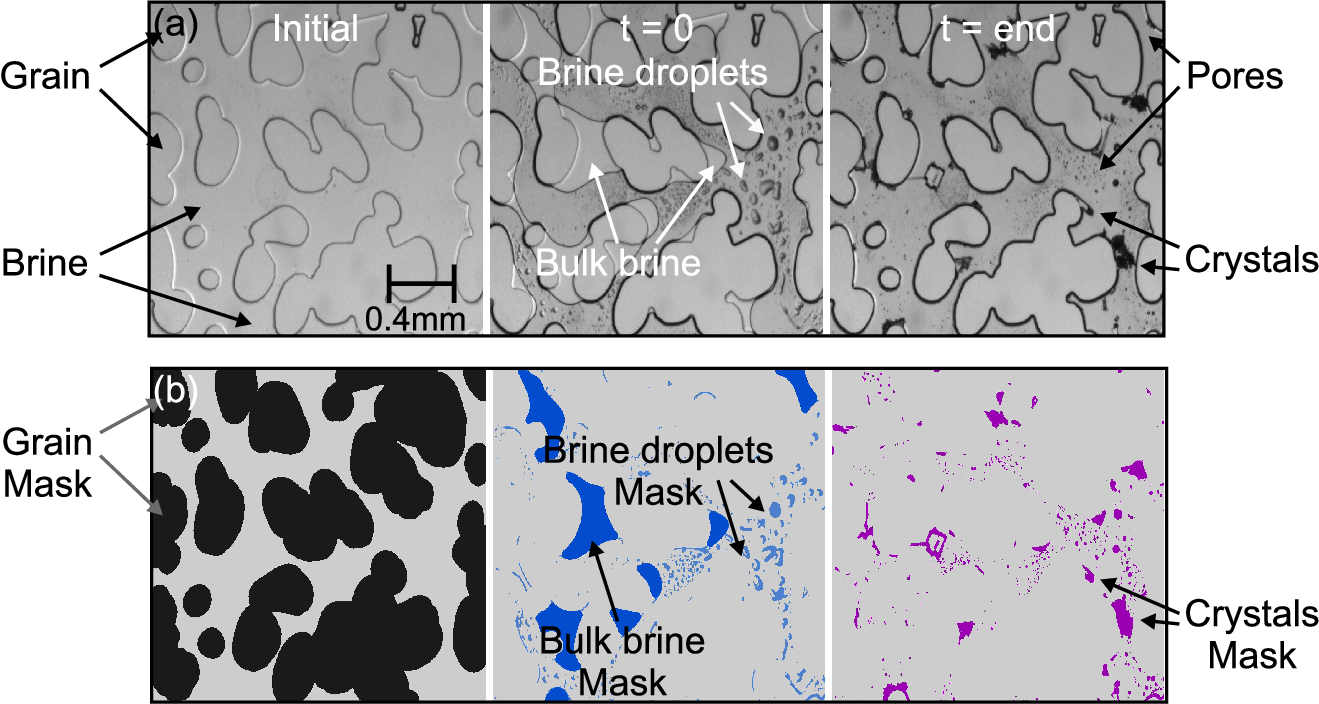}
\caption{(a) Microscopic images at a single chip position across experimental stages: (left) fully brine-saturated chip, (center) post-CO\textsubscript{2} breakthrough with partial brine displacement, (right) complete brine dry-out with salt crystals. Arrows indicate fluid and solid phases. (b) Image segmentation outputs: (left) pore network mask, (center) brine mask comprising bulk water and droplets, (right) final crystal distribution.}
\label{fig:figure2}
\end{figure}

We monitored the evaporation-precipitation dynamics within the microfluidic porous medium through a multistep experimental workflow, illustrated in Figure~\ref{fig:figure2}(a). Initially, the chip was evacuated under vacuum and saturated with the test fluids listed in Table~\ref{tab:IFT_Flow} using a high-pressure pump. Full saturation was confirmed by optical transparency because the refractive indices of the brine and glass substrate align, rendering the pore-glass boundaries faintly visible. Subsequently, the system was pressurized to 5 MPa with CO\textsubscript{2} through a separate dedicated inlet tube. To prevent residual brine from contaminating the gas stream, the brine injection line was disconnected and replaced with gas-specific tubing, avoiding valves or tee connectors that could retain liquid.

Following pressurization of the system, a predetermined pressure gradient imposed by the pump facilitated CO\textsubscript{2} breakthrough, initiating two-phase viscous displacement that displaced a portion of the brine. The subsequent evaporation phase began as residual brine within the pore network dissolved in the CO\textsubscript{2} stream, driven by the water solubility in the gas phase. Localized brine droplets persisted because of capillary trapping, with occasional detachment observed. As evaporation progressed and the liquid phase became concentrated, reaching the saturation limit, salt nucleation and crystallite formation marked the onset of detectable crystallization. Crystal growth and aggregation continued until complete dry-out, when no liquid or dissolved solute remained available.

After each experiment, the micromodel was cleaned by triple rinsing with distilled water, then vacuumed, and finally dried to ensure the pore network was restored for subsequent tests. Each fluid composition and flow rate condition was repeated three times.

Crystallization dynamics was quantified through high-resolution imaging and analysis. Initial brine saturation and final crystal coverage were assessed at 10 equidistant positions along the 20 mm length of the chip, while time-resolved evaporation and crystal growth were monitored at a fixed position 10 mm from the input (midpoint). A custom MATLAB-based image processing pipeline segmented the images (Figure~\ref{fig:figure2}(b)) to: (1) map the pore network in dry chips, defining porosity as the area fraction $\phi = A_p / A$ (where $A_p$ is the pore area and $A$ is the total area); (2) detect the distribution of the brine (bulk and droplets); and (3) track crystal evolution. The brine saturation ($S_w = A_w / A_p$) and the crystal coverage ($X_c = A_c / A_p$) were calculated, where $A_w$ and $A_c$ represent the brine and crystal areas, respectively, normalized to the pore area. This approach enabled a precise quantification of the kinetics of brine evaporation and crystal growth patterns under experimental conditions.

\section{Results and Discussion}
\subsection{Morphology and spatial distribution of brine saturation} 
	
	The fully saturated chip undergoes rapid desaturation during a CO\textsubscript{2} breakthrough. Part of the brine is immediately displaced from the chip. The other part is immobilized by the adhesion forces and is attached to the chip structure.  This point is considered to be the beginning of the brine evaporation process. However, brine flow can still be observed during evaporation. In this phase, brine drops can randomly detach from a chip grain upstream and attach to a grain downstream. This detachment process can continue for neighboring grains. The gas flow can also stretch the pool downstream. When the stretch reaches a glass grain, the brine pool can be further displaced because of the adhesion forces between the brine and the glass. Therefore, during the evaporation phase, some areas can undergo a rapid increase in saturation during attachment and a rapid decrease during brine detachment. To take these effects into account, the desaturation process should be measured for the largest number of brine pools. Therefore, in this study, images for 10 positions are being analyzed.
    
\begin{landscape}
		\begin{figure}
			\centering	
			\includegraphics[width=1.4\textwidth]{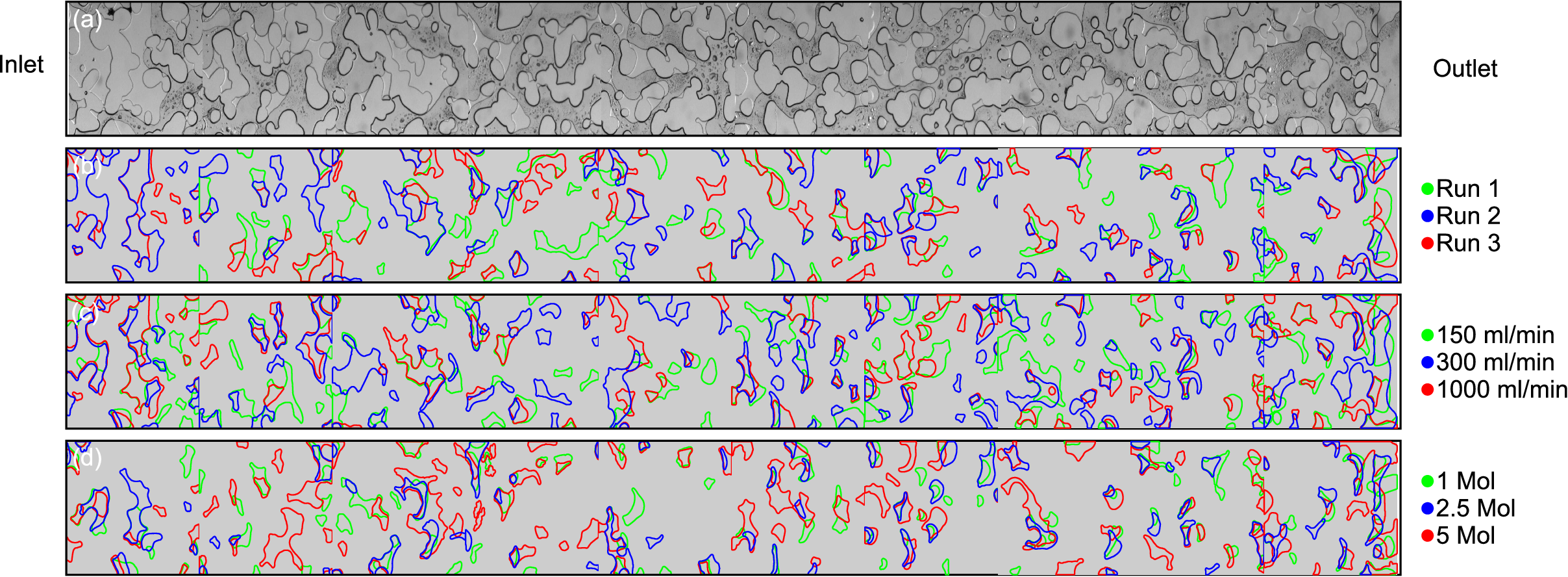}
			\caption{ Panoramic image of the chip along the flow direction. Brine distribution after initial displacement after CO\textsubscript{2} breakthrough (a) NaCl 5 mol/kg brine solution (W 5) flow rate 300 ml/min (b) three distinct experiments for fluid W5 and flow rate 300 ml/min. The brine boundaries of bulk water marked in colors for each experiment (c) chip image for three flow rates for W5 fluid with colors the bulk brine boundaries are marked (d) brine boundaries for three water NaCl solutions (W1, W2.5, W5 samples) and 1000 ml/min flow rate  }
			\label{fig:figure3}
		\end{figure} 
		
	\end{landscape}

    Figure~\ref{fig:figure3} shows panoramic images of a chip along a flow direction after CO\textsubscript{2} breakthrough. They cover a chip area of 20$\times$2 mm. The images are stacks of 10 positions measured with the laboratory setup. The joints of individual photos are visible as a shift in the perpendicular direction because of the imperfect alignment of the chip with respect to the XYZ motor moment. Figure~\ref{fig:figure3} (a) shows the residual brine in the chip for a displacement of 5 Mol NaCl brine and a CO\textsubscript{2} flow rate  of 300 ml/min. The right-hand side collecting channel is visible at the chip outlet. The inlet is not visible due to the chip tubing, which prevents the chip from moving to the relevant position. The left of the image is at a distance of 2 mm from the chip inlet. Various patterns of brine pools can be observed. In particular, brine occupies cavities in a glass. The brine surrounds the glass grains with a thin layer that widens on the downstream side of the glass patches. It also spans between adjacent glass grains, forming large pools. Large pools also form mainly on the downstream side of the glass grains. There are also visible drops inside the empty spaces that are remnants of the flowing fluid. The distribution of brine is not uniform along a chip. It strongly depends on the geometry of the glass patch. In particular, at the outlet, the brine occupies a large pore volume spreading between several glass grains. In the central part of the chip, individual pools are smaller and connect only single grains. That initial saturation of the brine $S_w$ directly affects the expected crystallization kinetics and the final coverage of the crystal. A thin layer of brine is expected to evaporate faster than a large pool. The precipitation of a single salt crystal in glass grains can trigger capillary flow and significantly accelerate the precipitation process \cite{dkabrowski2025surface}. Therefore, this saturation will be analyzed qualitatively and quantitatively depending on flow, position within the chip network, or type of fluid. 
	
	To demonstrate the stochastic nature of the breakthrough process and the resulting residual brine saturation distribution, Figure~\ref{fig:figure3} (b) presents the initial brine distribution immediately after CO\textsubscript{2} breakthrough for three consecutive measurements for the displacement of 5 mol / kg of NaCl brine (W5) and a CO\textsubscript{2} flow rate of 300 ml/min. The colored outlines indicate the boundaries of the brine pools identified from the bulk brine masks in each run. For better visualization, water in the form of a droplet is excluded. 
    
    In Run 1, no sharp image was recorded for the first portion; therefore, no green signal was present. It can be seen that, despite the same measurement conditions, the brine pools differ for each run. There are positions in which pools are formed for each run. In particular, small pools form in the cavities of glass grains. In contrast, large pools vary significantly in size, shape, and location. This shows that during the breakthrough, the CO\textsubscript{2} flow paths evolve stochastically, following random routes through the channel network. These complex invasion sequences are affected by local pore-scale features and give rise to hysteresis; meaning that every time the experiment is repeated in the same medium, the brine distribution after breakthrough is different. This kind of random path creation has been previously analyzed for the oil-brine system and has shown that channel geometry, IFT, viscosity, and wettability have a major impact on breakthrough dynamics \cite{gogoi2019review,schumi2020alkali,saadat2020development}. 
    
    Figure~\ref{fig:figure3} (c) shows the brine distribution for the displacement of 5 mol / kg of NaCl brine (W5) at three different flow rates: 150, 300 and 1000 ml / min. Since the brine distribution is random, the position of the pools does not show a consistent dependence on flow rate. However, total brine saturation can still be influenced by flow conditions. In this case, no systematic trend is observed. Large pools form regardless of the flow rate. A quantitative analysis of the effect of flow on saturation will be presented later in the article.
	
	IFT can affect a breakthrough process on a microchip \cite{gogoi2019review}. Therefore, in this study, several fluids are used in chip flooding experiments. Figure~\ref{fig:figure3} (d) shows an exemplary distribution of brine pools for three brine samples with a concentration of 1, 2.5 and 5 mol / kg NaCl and a flow rate of 1000 ml/min. An increase in salinity results in a higher IFT table~\ref{tab:IFT_Flow}. However, no evident impact of salinity on the brine distribution can be observed in the microscopic image. Further statistical analysis of brine saturation will be conducted in the later part of the manuscript. The initial saturation images obtained for all the fluids and flows shown in Table~\ref{tab:IFT_Flow} can be quantitatively analyzed to determine the influence of the flow and fluid characteristics on the initial saturation of the brine $S_w$.
	
	\begin{figure}
		\includegraphics[width= 0.95\textwidth]{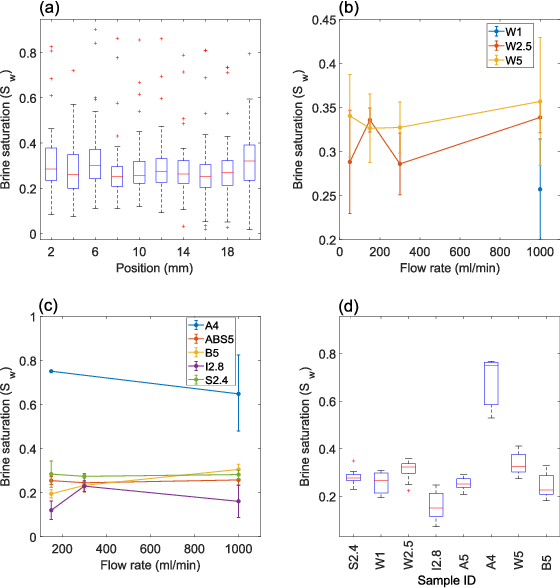}
		\caption{Distribution of initial brine saturation $S_w$ (after initial CO\textsubscript{2} breakthrough) measured by microfluidic setup with respect to (a) position within form the chip inlet. (b-c) flow rate. Labels indicates sample ID (d) fluid sample (ID are describe in the Table~\ref{tab:IFT_Flow}. Red mark denotes the median, box indicate the 25th and 75th percentiles while the whiskers extreme data points not considered outliers. Symbol $+$ plots the outliers.}
		\label{fig:figure4}
	\end{figure} 
	
	Fig.~\ref{fig:figure4} (a) shows brine saturation $S_w$ with respect to the position of the chip inlet. Analysis of the median value shows that the saturation varies significantly between positions. The lowest values are for positions 8 and 16 mm, while the highest values are at 6 mm and 20 mm. It shows that on average, saturation is affected by the pore-space geometry. However, $S_w$ is not correlated with position; in particular, it does not depend on pressure (which decreases downstream). For different fluids and runs, the saturation $S_w$ changes from 0.1 to 0.8 at the given position. Therefore, the variations are random, which is a result of the random nature of a breakthrough. Outliers are most notably ammonia solutions samples.
    
    Figures~\ref{fig:figure4} (b-c) show saturation with respect to flow rate,  For each experiment, the data have been averaged over positions. Figure~\ref{fig:figure4} (b) shows the relation between $S_w$ and the flow rate for fluids W1, W2.5, and W5, while Fig. ~\ref{fig:figure4} (c) shows $S_w$ for fluids A4, ABS5, I2.8 and S2.4. It shows that the initial saturation in the micromodels used considered flow rates is not significantly affected by the flow rate. It should be pointed out that for W1 only the experiment with 1000 ml/min flow rate has been performed. In the case of fluid A4 due to chip clogging, no data have been obtained for 300 ml / min. This shows that for the flow rates considered, the CO\textsubscript{2} stream does not significantly displace the brine. Brine that is attached to the glass grains after CO\textsubscript{2} breakthrough. Therefore, reducing initial saturation with a modification of the injection flow rate is not feasible. It should be pointed out that for A4 fluid and 300 ml/min flow rate and partially for 150 ml/min flow no $S_w$ have been obtained due to chip clogging and imperfect image segmentation method.
    
    Similarly, NaCl concentration (Fig.~\ref{fig:figure4} (d)) shows the impact of NaCl concentration on $S_w$. For water-based solutions, $S_w$ does not change significantly with concentrations. The median saturation is 0.27-0.28 for solutions of 1, 2.4 and 5 M. However, for a 2.5 M it is 0.32. For 2.8 M and 4 M solutions, $S_w$ is 0.15 and 0.75 respectively.  Similarly, Figure~\ref{fig:figure4} (d) shows brine saturation for a given fluid. The lowest median saturation observed for an I2.8 fluid (solutions with propan$-2-$ol), shows that IFT impacts significantly $S_w$. The lower IFT results are the lower saturation as a result of the higher mobility of the fluid compared to that of the pure water+NaCl system. In contrast, the highest saturation for A4 (ammonia solutions) shows that for each experiment with ammonia, rapid crystallization of ammonia bicarbonate results in brine trapping and increases $S_w$ compared to water + NaCl brine, although its IFT is similar, which will be described in detail in the later part of the manuscript. Comparison of W5 and ABS5 fluid shows that the addition of alkylbenzene sulfonate decreases $S_w$. Similarly, the base solutions B5 result in significantly lower IFT and $S_w$ compared to W5.
    
	\subsection{Dynamics of brine evaporation and halite precipitation}
	
	After CO\textsubscript{2} breakthrough brine pools remain generally stable. CO\textsubscript{2} flow evaporates brine. As a result, the pools become smaller. When the brine in a given pool reaches the saturation limit, crystal nucleation occurs. When evaporation progresses, the crystals grow. After a certain time, the brine evaporates completely and the process stops, and no further change in the crystal pattern is observed. The following analysis focuses on determining how the flow rate or fluid composition impacts both the time of the first crystal nucleation and the total time of the crystallization process.
	
	\begin{figure}
		\includegraphics[width=0.95\textwidth]{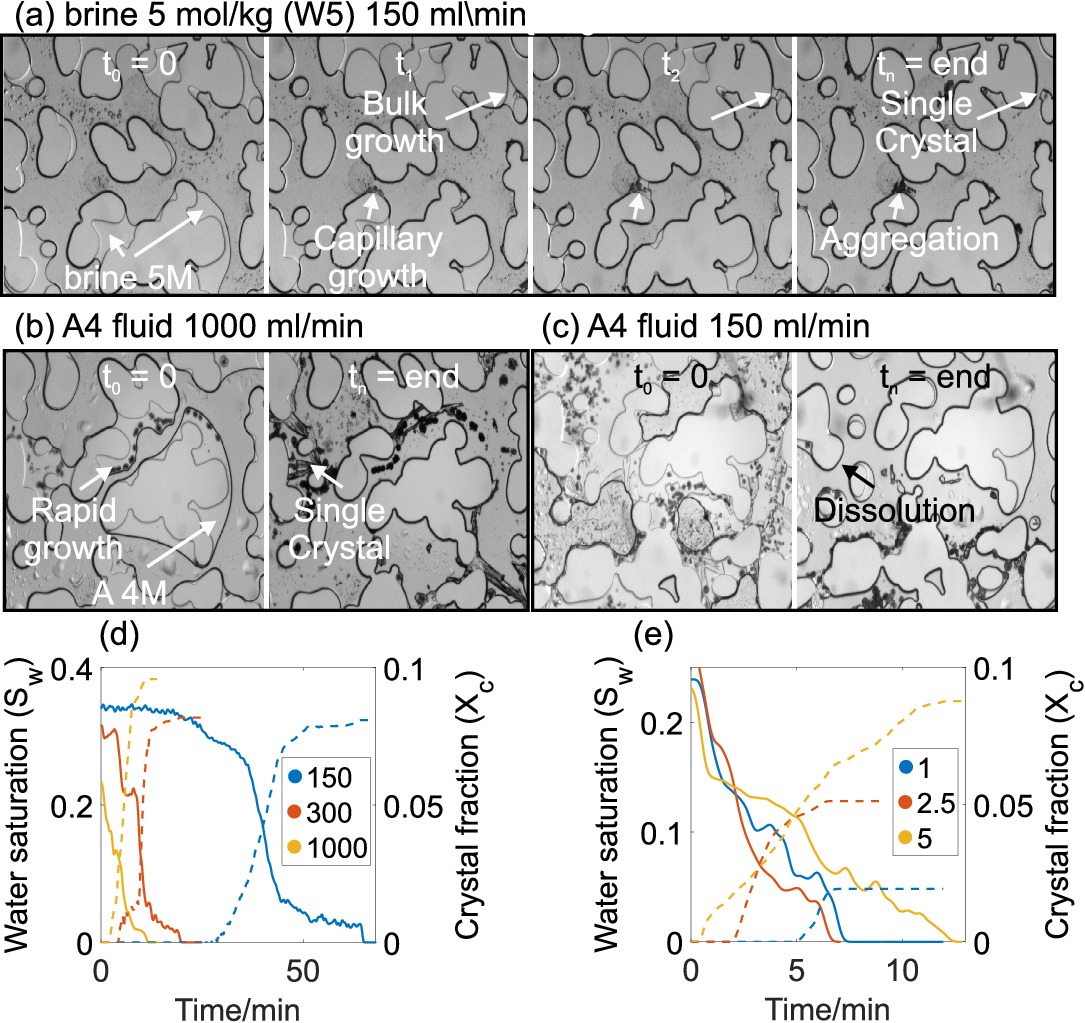}
		\caption{Crystallization and dissolution dynamics under different conditions.
			(a) Growth process in a 5M brine solution (W5 fluid) 150 ml/min flow rate. Bulk and capillary growth leading to single crystal formation and aggregation (exemplary crystallines marked with arrows)
			(b) Rapid crystal growth in A4 fluid (ammonia solution) 1000 ml/min flow rate after initial CO\textsubscript{2} breakthrough (left panel), final crystal coverage at the end of fluid dryout (right panel)
			(c)  Rapid crystal growth in A4 fluid (ammonia solution) 150 ml/min flow rate after initial CO\textsubscript{2} breakthrough (left panel), crystal
			dissolution and growth after chip clogging (right panel) 
			(d) Exemplary time-dependent evolution of water saturation ($S_w$- solid line) and crystal fraction ($X_c$- dashed line)
			for different flow rates (150, 300, 1000 ml/min), fluid W5
			(e) Exemplary time-dependent evolution of water saturation ($S_w$) and crystal fraction ($X_c$)
			for different NaCl concentrations  (1M, 2.5M, 5M), fluid W5 1000 ml/min flow rate, highlighting the interplay of drainage, precipitation, and evaporation in three stages }
		\label{fig:figure5}
	\end{figure}

	Figure~\ref{fig:figure5} shows an example of the temporal evolution of the crystallization process in the microfluidic chip. Fig.~\ref{fig:figure5}(a) shows four distinct time points. First brine saturation after CO\textsubscript{2} breakthrough. This point is considered to be the beginning of an evaporation process. The arrow shows the bulk brine residuals which undergo evaporation. It can be seen that large brine pools extended through neighboring glass patches. Brine also fills concave cavities in glass patches and forms a film at the convex surfaces on the downstream side of the glass patches. Because glass is hydrophilic, a liquid film forms on the glass surfaces and pools marked with white arrows can be connected. However, the current resolution of the system does not allow visualization of the thin-liquid films or capillary flow in the film. The second and third panels show partial dry-out when the crystals are in different stages of growth. First, brine tends to evaporate from a film on convex surfaces. Evaporation from concave cavities occurs much more slowly. When a large pool, which is adjacent to several glass patches, evaporates, it can separate into distinct brine pools. As a result, the connectivity of the system decreases and further reduces the mobility of the brine.

    With white arrows, two exemplary types of crystal growth are shown. The first type, crystal, grows inside the brine pool (bulk growth) and is shown on the top right side of the panel. The second type, crystal, grows outside a brine pool (aggregation). In this case, brine is transported by capillary forces from a nearby brine pool that feeds crystal growth.  When brine connectivity is high, this process can result in large crystal growth. For a reservoir with unlimited access to brine, it is the main clogging mechanism. The right panel shows the final crystal coverage after complete brine dry-out. Crystals tend to form on the convex surfaces of glass patches. Films on convex surfaces evaporate faster. Therefore, in those regions, crystals first form. That crystals promote brine evaporation. The time point at which the first crystal appears is the beginning of crystallization. The time when the last brine evaporated is considered the end of the crystallization and evaporation process.

    Figure~\ref{fig:figure5} (b,c) shows crystallization in an ammonia solution (A4 fluid).  fluid shows a different behavior compared to other solutions.  In addition to the residual brine, ammonia bicarbonate crystals form inside brine pools and droplets right after the breakthrough. They grow in the whole area of the chip; in particular, close to the phase boundary (brine CO\textsubscript{2}). This compound is formed in a chemical reaction of carbonic acid $\text{HCO}_{\text{3}}$ with ammonia in the reaction \cite{sutter2017solubility}:
\begin{equation}
    \text{NH}_{\text{3}}+\text{HCO}_{\text{3}}^{-} \rightleftharpoons \text{NH}_{\text{2}}\text{COO}^{-}
\end{equation}

   Crystals that are hydrophilic act as a frame for a brine.  As a result, for the A4 fluid, the initial saturation is higher than for the other fluids. The brine is adjoined to many glass grains, resulting in high connectivity of the system i.e. the brine can freely flow in inside a pool. However, extensive growth of ammonia bicarbonate crystals causes chip clogging or makes it impossible to effectively segment images.

    The left panel of Fig.~\ref{fig:figure5} (b) shows that the ammonia bicarbonate crystals form right after the CO\textsubscript{2} breakthrough. This is in contrast to water-based solutions, where crystallization starts after partial evaporation of the brine. Those crystals trap water, which results in a higher initial $S_w$ for ammonia compared to other fluids. Fig.~\ref{fig:figure5} (b) (right panel) shows the final crystal coverage of the chip space. Crystallization from water and an ammonia solution can result in the formation of long single crystals. In contrast to W5 brine, where the crystals form a polycrystalline aggregation of small square single crystals. The current system cannot distinguish between NaCl crystals and ammonia bicarbonate crystals, giving only the final crystal coverage of the pore space.

    Fig.~\ref{fig:figure5} (c) shows the exemplary case when the chip became clogged during crystallization. Clogging is a result of crystal growth in chip-tubing connections, which is reflected in the reducing or completely stopping the flow regardless of increasing a pressure gradient in the chip. The flow drop decreases or prevents the evaporation of the brine. Resulting in a smaller crystal coverage. Moreover, in part of a chip crystals present at the begging if the processes (marked with arrow) can be dissolved. This leads to the conclusion that, to some extent, ammonia bicarbonate, which precipitates from a water-ammonia solution, leaves an undersaturated brine that can flow towards the chip outlet, where it can dissolve previously formed crystallites. Due to the high $S_w$ and therefore high connectivity of the glass grains, the flow is not driven by CO\textsubscript{2} flow but is dominated by diffusion. In particular, it can occur on the clogged chip.

    Figure~\ref{fig:figure5} (d) shows an exemplary time profile of water saturation ($S_w$ - solid lines) and crystal fraction ($X_c$- dashed lines) time profile for different flow rates and a fluid W5 (water+NaCl 5 mol/kg brine). The color label denotes flow rates of 150, 300 and 1000 ml / min. An increase in flow rate results in significantly faster brine evaporation. For high flow rates, crystals nucleate earlier than for lower flow rates. Moreover, total crystal formation progresses faster and ends earlier. Similarly, Figure~\ref{fig:figure5} (e) shows exemplary time profiles of water saturation and crystal fraction for three NaCl concentrations in (W1, W2.5 and W5 mol / kg) and a flow rate of 1000 ml / min. Solid and dashed lines denote $S_w$ and $X_c$, respectively. It can be seen that a lower concentration crystallization occurs later. This is because more brine has to evaporate to reach the saturation limit. Moreover, a higher NaCl concentration results in a higher final $X_c$. 
	
	\begin{figure}
		\includegraphics[width=0.95\textwidth]{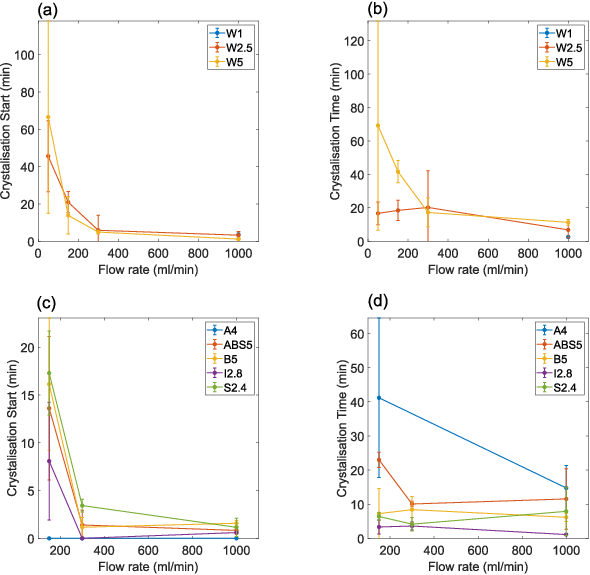}
		\caption{Impact of flow rate and sample type on crystallization start time and total crystallization duration with respect to flow rate (a-b) crystallization start time and the total crystallization duration time for fluids W1, W2.5 and W5 (c-d) crystallization start time and the total crystallization duration time for fluids A4, ABS5, B5, I2.8 and S2.4 }
		\label{fig:figure7}
	\end{figure}     			

    Crystal growth can be described by two time points. First, duration between breakthrough and first crystal nucleation. Second, the duration of a crystallization is the time from the first nucleation to complete dry-out. These two duration times can be analyzed for the fluid and flow rates considered.  
    
    Figure~\ref{fig:figure7} shows the impact of flow rate and fluid type on the start time of crystallization (time from the CO\textsubscript{2} breakthrough to the first crystal nucleation) and the duration of crystallization (time from the first crystal nucleation to the complete evaporation of the brine). Figure~\ref{fig:figure7} (a,c) shows that a faster flow rate results in faster nucleation. Is connected to faster brine evaporation. Moreover, it enhanced the mixing of a brine in a pool, reducing concentration gradients and promoting more uniform supersaturation, thereby accelerating nucleation. Similarly, a higher flow rate shortens the total crystallization duration time (Fig.~\ref{fig:figure7} (b,d)). Moreover, higher flow rates decrease duration variations. Crystallization begins faster for more volatile samples, in particular, ammonia + water (A4) solution and alcohol (I2.8) solution. However, (W5, ABS 5) have delayed nucleation with high variability (Fig.~\ref{fig:figure7} (c)). 
    
    Figure~\ref{fig:figure7} (b,d) shows that the crystallization duration time is longer for higher NaCl concentration. For alcohol solutions (I2.8), crystallization is faster due to high volatility, while for ammonia (A4) crystallization it is long due to high $S_w$, water trapping in the crystal structure. Brine with KOH hydroxide significantly lower crystallization duration compared to W5 or ABS5 fluid. It can be associated with changing pH, which can reduce the solubility of NaCl or K$^+$ and OH$^-$ ions, and they may be additional nucleation promoters that enhance ion pairing and aggregation.
	
	%%%%%%%%%%%%%%%%%%%%%%%%%%%%%%%%%%%%%%%%%%%%%%%%%
	\subsection{Morphology and spatial distribution of salt crystals}
	
	Brine evaporation triggers crystal growth. The analysis of the temporal crystallization dynamics shown in detail in Figure~\ref{fig:figure5} shows that the crystals grow within the brine pools and outside those pools. In this case, crystal growth is fed by the capillary flow of the brine from neighboring brine pools. Crystals can form single crystals (in particular for ammonia solution) but most notably form polycrystalline aggregates.
    
\begin{landscape}
		\begin{figure}
			\includegraphics[width=1.4\textwidth]{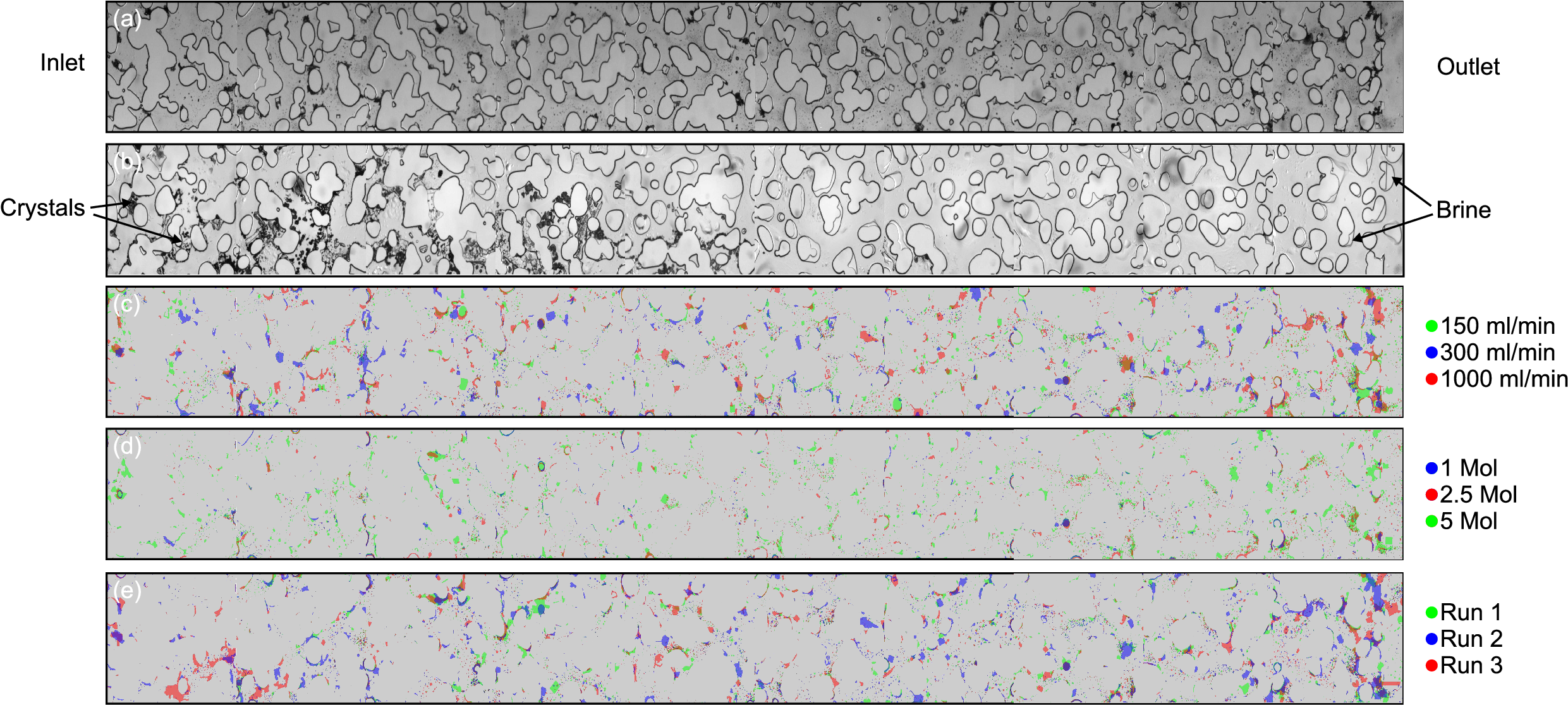}
			\caption{ Panoramic image of the chip along the flow direction. Crystal coverage after complete water dry-out  (a) NaCl 5 mol/kg brine solution (W 5) flow rate 300 ml/min (b) image for a single experiment after complete chip clogging for ammonia solution (A4) and initial flow rate of 150 ml / min. At the inlet chip, the brine evaporated, and only the crystals are visible downstream; more bulk water is visible. At the outlet, the crystals were dissolved (see Fig.~\ref{fig:figure3} (e)) (c) three different experiments for the W5 fluid and a flow rate of 300 ml / min. The colored crystallites are marked for each experiment (d) chip image for three flow rates for the W5 fluid with colored crystalline is marked (e) crystal coverage for three water NaCl solutions (W1, W2.5, W5 samples) and the 1000 ml / min flow rate }
			\label{fig:figure8}
		\end{figure} 
	\end{landscape}

    Figure~\ref{fig:figure8} shows panoramic images of a chip along a flow direction after complete brine evaporation. The final crystal distributions are observed for the same experiments as shown for the initial brine saturation in Figure~\ref{fig:figure4}. The images are a stack of images taken in 10 positions and covering a chip area of 20$\times$2 mm. Figure~\ref{fig:figure8} (a) shows the crystal distribution in the chip for a 5 Mol NaCl brine displacement and a CO\textsubscript{2} flow rate 300 ml/min. From the image a large crystal can be seen in contact with glass patches. In the void area, only small crystals are observed. Moreover, they are not uniformly distributed. Some glass grains are free from crystallites or host only small-sized crystallites. However, other hosts have multiple large aggregations. Most notably, they form in large numbers close to the chip outlet in comparison to those in the middle positions. This is due to the presence of a collecting channel that has a significant volume in relation to the porous structure. To determine whether the final position distribution of the crystal is random, a chip dry-out experiment was repeated.

	%%%%%%%%%%%%%%%%%%%%%%%%%%%%%%%%%%%%%%%%%%%%%%%%%
	
Figure~\ref{fig:figure8} (b) shows an example of an initial flow rate of 150 ml / min when the chip became clogged and no flow was possible for a certain time. In the case of initial brine saturation shown in the Figure~\ref{fig:figure5} (c) ammonia bicarbonate crystals can form after CO\textsubscript{2} breakthrough inside brine pools. 
 In the case of a clogged chip, it can be observed that the brine evaporates close to the chip inlet. As a result, NaCl crystals formed large aggregates with ammonia bicarbonate crystals. The final crystal coverage of aggregates and single crystals is significantly higher than $X_c$ for water-based solutions (despite a lower total NaCl concentration than for the W5 fluid). 
     
  In contrast, close to the chip outlet, a completely different process is observed. The initial ammonia bicarbonate crystals are dissolved. Only partial evaporation of the brine pools occurred where both the crystals and fluid are visible. However, near the chip outlet crystals are not visible. The ammonia bicarbonate crystals were dissolved and the NaCl crystal did not precipitate. The process has been shown in detail in a temporal image (Figure~\ref{fig:figure5} (c)).
    
	 Figure~\ref{fig:figure8} (c) shows the crystal distribution for three consecutive measurements for 5 mol/kg NaCl evaporation and a CO\textsubscript{2} flow rate of 300 ml / min. In colors, crystallites formed for consecutive experiments are shown. From this panel it can be concluded that crystallites do not have the same spatial distribution for each experiment they form in random positions. At some positions there are significantly more crystals for different experiments. For example, for Run 3 (red) the crystals formed in large numbers at the chip input. In the opponents for Run 1 (green), they form mostly in the middle of the chip. For each Run, there is a high concentration of crystallites at the chip outlet. That random positions of the crystalline are the result of a random distribution of a brine pool after the CO\textsubscript{2} breakthrough and the probabilistic nature of the nucleation process.

     To determine the impact of the flow rate on the crystal distribution and the crystal coverage $X_c$, dry-out experiments have been performed for various flow rates. Figure~\ref{fig:figure8} (d) shows the exemplary crystal positions after evaporation of the 5 mol/kg NaCl brine at three CO\textsubscript{2} flow rates of 150, 300, and 1000 ml / min. Similarly to the previous case, crystals form in random positions, and no evident impact of flow rate is visible on $X_c$. In contrast, Figure~\ref{fig:figure8} (e) shows the crystal distribution after evaporation of three brine concentrators 1, 2.5 and 5 mol/kg NaCl (fluids W1, W2.5 and W5). There is significantly lower crystal coverage for lower NaCl concentrations than for higher ones. Moreover, for highest salinity, the crystal aggregates are larger than those for lower concentrations. 
		
	\begin{figure}
		\includegraphics[width=0.95\textwidth]{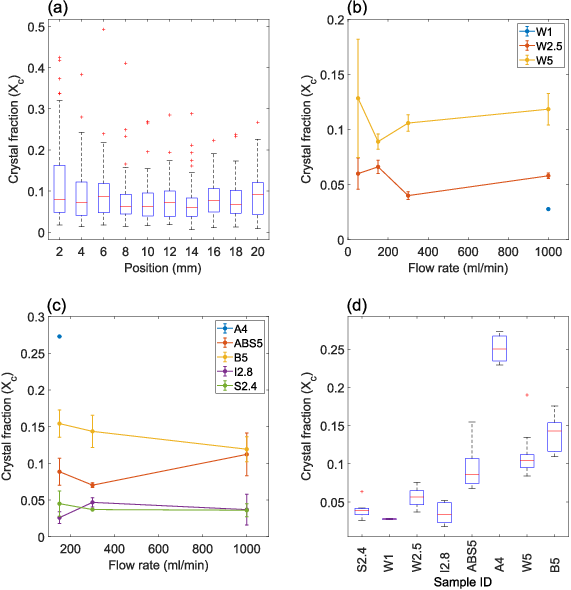}
		\caption{Distribution of final crystal fraction measured by microfluidic setup with respect to (a) position within form the chip inlet. (b-c) flow rate samples ID are shown in the label (d) fluid sample (ID are describe in the Table~\ref{tab:IFT_Flow}. Red mark denotes the median, box indicate the 25th and 75th percentiles while the whiskers extreme data points not considered outliers. Symbol $+$ plots the outliers form A4 fluids. }
		\label{fig:figure9}
	\end{figure}     	
    
	The final crystal coverage can be analyzed for each measurement and position. For the ammonia solution (A4 fluid), only results for fully dried positions are considered. There are only data for 150 ml/min. Figure~\ref{fig:figure9} shows the statistic of the measured crystal fraction after complete dryout (void space coverage). Fig.~\ref{fig:figure9} (a) shows the distribution with respect to the different positions of the chip input. The median of distribution (red line) shows the crystal fraction. It can be seen that the highest fraction is at the chip outlet, which is connected with crystal aggregation at the outlet collecting channel. The analysis of variance (ANOVA test and Spearman correlation parameter analysis) showed a lack of correlation between the crystal fraction and position. It also showed significant variations between positions. The variations are purely due to the randomness of the crystallization process. 
    
    Figures~\ref{fig:figure9} (b-c) show the final crystal coverage with respect to the flow rate. For each experiment, data were averaged over positions. For W1 experiment has been conducted only for the flow rate 1000 ml/min. Fig.~\ref{fig:figure9} (b,c) shows that in the following micromodel for the flow rates considered the crystal fraction does not depend significantly on that flow rate. Since adhesion forces dominate over kinetic forces, the CO\textsubscript{2} stream does not alter the saturation of the brine and the capillary-driven flow, which contributes to the total crystal fraction. Fluids with higher NaCl concentrations (Fig.~\ref{fig:figure9} (b,c))  promote greater crystal formation, which is consistent with typical crystallization behavior, where supersaturation enhances nucleation and growth. The increase in variance at high concentrations suggests that some regions may experience localized precipitation, leading to heterogeneous crystallization.  In general, the trend can be considered linear. 
    
    In Figure~\ref{fig:figure9} (d) one can observe that a higher NaCl concentration in a brine is associated with a higher crystal fraction. However, in the case of the I2.8 fluid (solutions with propan$-2-$ ol), the crystal coverage is significantly decreed. The lower IFT and the higher volatility are associated with a lower water saturation $S_w$ shown in Figure~\ref{fig:figure4} (d) and, as a result, decreases $X_c$. In particular, the I2.8 fluid has significantly lower $X_c$ than the W2.5 fluid. In contrast, for ABS5 fluid the decrease in crystal fraction is only moderate compared to that of W5 fluid. Therefore, the final crystal fraction is affected for both the IFT and the volatility alternation. S2.4 fluid shows a lower crystal fraction than the W2.5 fluid, which can be connected with the exclusion of outliers (red dot) from the median. 
    
    B5 fluid has a higher crystal fraction than W5 in addition to the same NaCl concentration. In this case, the K$^+$ and OH$^-$ ions not only enhance the crystallization speed but also increase the salt aggregation. 
    
     For a fluid $4$ (ammonia solutions), a high $X_c$ is observed. This is directly related to a higher brine saturation. The brine saturation is high because of the rapid crystallization of ammonia bicarbonate and fluid trapping, which corresponds to high crystal coverage and as a result of flow clogging. It should be stressed that for A4 fluid crystal coverage is higher than for W5 even in solution the total mass of NaCl is lower. 

     The main factor limiting crystal growth is the limited access to brine in the micromodel. In a real reservoir, the brine can be continuously supplied from areas far from the well.  
	
	\section{Conclusions}
The microfluidic experiments conducted in this study offer valuable insight into the kinetics of salt precipitation under CO\textsubscript{2} injection conditions. Higher CO\textsubscript{2} flow rates have been shown to lead to faster brine evaporation and earlier begging of crystal nucleation. However, within the parameter space explored and the 2D micromodel configuration, no significant impact of the flow rate has been observed on initial brine saturation $S_w$ or final crystal fraction $X_c$. The geometry of residual brine and its direct influence on the morphology and location of the crystal is random. As such, the experiments do not indicate a clear potential for using the flow rate alone to mitigate clogging or injectivity loss caused by salt precipitation in subsurface formations. It is important to note that while microfluidic tests reproduce key pore-scale processes, they cannot fully replicate the complexities of real reservoir systems - including heterogeneity, 3D connectivity, access to brine source and geological variability. Therefore, caution should be exercised when extrapolating these results directly to large-scale CO\textsubscript{2} storage scenarios. However, these findings contribute to a mechanistic understanding of the dynamics of salt precipitation and can inform the development of improved models or help refine injectivity risk assessments. The future work could cover a wider range of flow rates, 3D geometries, or alternative geometry of the micro-model, e.g. a thicker structure or smaller pores. 

Studies directly showed that the capillary-driven flow of brine between neighboring pools plays a significant role in crystal formation. The crystal growth patterns, both within and outside the brine pools, indicate that crystallization could be mitigated by altering fluid properties, especially those that impact supersaturation and evaporation rates.

Analysis showed that higher NaCl concentrations promote more crystallization, but this also leads to increased variability and potential localized precipitation, which could cause uneven clogging. Modifying the concentration of brine could influence the deposition of salts, because higher concentrations generally lead to more salt formation, exacerbating clogging risks. Ionic additives such as K$^+$ and OH$^-$ lead to faster crystal formation and higher final crystal fraction. In rock porous systems, it can be particularly important when carbonic acid can react with particular minerals.

Both volatility and surface tension have an impact on crystallization. Decreasing IFT and an increase in the volatility of the fluid by surfactant additives or alcohol solutions can potentially reduce crystallization and clogging. 

Ammonia solutions (A4)  are characterized by a higher initial water saturation, due to the rapid crystallization of ammonia bicarbonate and, as a result, a higher crystal fraction. This leads to a higher risk of clogging, as the crystal formation is more extensive. This shows that in multicomponent brine / rock systems, not only NaCl precipitation can occur. For an A4 fluid, the ammonia bicarbonate crystals grow immediately due to the dissolution of CO\textsubscript{2} in the solution. No brine evaporation is required. This can potentially lead to rapid clogging. Therefore, a detailed understanding of the mineralization of the rock and brine is required. So, it is possible to model all potential crystallization or precipitation reactions.  

	\section*{Data availability}
	Data supporting the findings of this study are publicly available in the RODBUK open repository via DOI: \href{https://doi.org/10.58032/AGH/31YOBC}.
	
	\section*{Acknowledgments}						
	This work was supported by the project 'solid and salt precipitation kinetics during CO\textsubscript{2} injection into the reservoir', funded by Norway Grants (Norwegian Financial Mechanism 2014-2021) under grant number UMO-2019/34/H/ST10/00564. This research was also partially funded by the EEA and Norway grants under grant number NOR / PONORCCS / Agastor / 0008/2019-00.
	
	\bibliography{mybibfile}
	%	\section*{References}

\end{document}